\documentclass[11pt]{article}
\usepackage[margin=0.85in]{geometry}
\usepackage[T1]{fontenc}
\usepackage{amsmath}
\usepackage{textcomp}
\usepackage[round]{natbib}
\usepackage{hyperref}
\usepackage{xurl}
\usepackage{booktabs}
\usepackage{microtype}
\usepackage{xcolor}
\usepackage{enumitem}
\usepackage{tikz}
\usepackage{array}
\usepackage{authblk}
\newcommand{\FlipBlockTable}{%
\begin{tabular}{rlccrrrrrrrr}
\toprule
Block & Site & Wording & Kind & $c$ & $L$ & $H$ & High & Flipped & Above $H$ & FP low & FP high \\
\midrule
1 & North & A & A & 8 & 14 & 42 & 16 & 10 & 44 & 7 & 19 \\
2 & East & B & A & 9 & 24 & 58 & 47 & 17 & 60 & 12 & 39 \\
3 & West & A & B & 30 & 17 & 63 & 57 & 21 & 66 & 8 & 32 \\
4 & North & B & B & 18 & 13 & 55 & 37 & 16 & 65 & 2 & 23 \\
5 & North & A & A & 5 & 13 & 50 & 40 & 10 & 58 & 10 & 18 \\
6 & East & B & A & 13 & 20 & 37 & 28 & 15 & 46 & 15 & 24 \\
7 & North & A & B & 18 & 14 & 53 & 32 & 16 & 63 & 9 & 18 \\
8 & East & B & B & 30 & 15 & 51 & 49 & 28 & 55 & 2 & 21 \\
9 & North & A & A & 6 & 11 & 26 & 18 & 9 & 28 & 2 & 14 \\
10 & East & B & A & 5 & 12 & 59 & 34 & 7 & 67 & 9 & 25 \\
11 & West & A & B & 21 & 11 & 53 & 51 & 15 & 63 & 3 & 21 \\
12 & North & B & B & 19 & 13 & 43 & 22 & 17 & 53 & 11 & 19 \\
13 & North & A & A & 8 & 17 & 61 & 43 & 10 & 64 & 8 & 23 \\
14 & North & B & A & 10 & 18 & 59 & 53 & 14 & 68 & 5 & 33 \\
15 & East & A & B & 15 & 8 & 27 & 19 & 12 & 31 & 5 & 10 \\
16 & East & B & B & 28 & 16 & 64 & 35 & 22 & 68 & 9 & 38 \\
17 & West & A & A & 6 & 10 & 26 & 12 & 8 & 29 & 2 & 13 \\
18 & East & B & A & 8 & 18 & 36 & 26 & 12 & 44 & 15 & 23 \\
19 & East & A & B & 27 & 17 & 50 & 32 & 19 & 57 & 7 & 22 \\
20 & East & B & B & 18 & 9 & 47 & 28 & 12 & 55 & 7 & 17 \\
21 & West & A & A & 7 & 12 & 24 & 19 & 10 & 30 & 2 & 14 \\
22 & West & B & A & 10 & 14 & 32 & 19 & 12 & 36 & 10 & 21 \\
23 & North & A & B & 30 & 12 & 56 & 39 & 19 & 66 & 3 & 30 \\
24 & North & B & B & 27 & 18 & 63 & 34 & 23 & 70 & 7 & 36 \\
25 & East & A & A & 14 & 25 & 45 & 32 & 23 & 51 & 19 & 31 \\
26 & East & B & A & 10 & 22 & 48 & 39 & 13 & 56 & 20 & 33 \\
27 & North & A & B & 31 & 18 & 62 & 39 & 24 & 70 & 8 & 20 \\
28 & East & B & B & 25 & 10 & 58 & 53 & 13 & 64 & 6 & 14 \\
29 & East & A & A & 9 & 16 & 64 & 51 & 12 & 67 & 12 & 29 \\
30 & East & B & A & 7 & 11 & 44 & 16 & 9 & 54 & 4 & 24 \\
31 & West & A & B & 16 & 9 & 44 & 40 & 12 & 52 & 3 & 18 \\
32 & East & B & B & 25 & 13 & 46 & 36 & 15 & 48 & 4 & 18 \\
\bottomrule
\end{tabular}}
\newcommand{\DomainTable}{%
\begin{tabular}{rllll}
\toprule
Dom. & Domain & Unit (service) & Adapted part & Embedded in \\
\midrule
1 & Aquarium & seawater treatment & filter interface collars & tank walls \\
2 & Athletics Center & performance timing & timing-system anchors & running lanes \\
3 & Botanical Garden & mist irrigation & spray-head connectors & greenhouse frames \\
4 & Cable Car & cable inspection & inspection-head sockets & tower crossbeams \\
5 & Campground & utility servicing & utility connection plates & camping pads \\
6 & Concert Hall & stage-lighting service & light fixture sockets & ceiling trusses \\
7 & Convention Center & visitor assistance & information kiosk docks & foyer columns \\
8 & Cruise Terminal & boarding service & gangway alignment sockets & quay walls \\
9 & Equestrian Center & horse rehabilitation & gait sensor plates & exercise tracks \\
10 & Ferry Port & vessel charging & charging interface plates & berth foundations \\
11 & Film Studio & motion capture & tracking beacon docks & soundstage walls \\
12 & Funeral Park & grounds irrigation & irrigation-control couplers & buried conduits \\
13 & Golf Course & turf monitoring & soil-probe sleeves & fairways \\
14 & Ice Rink & ice condition monitoring & temperature probe sleeves & rink floors \\
15 & Marina & berth servicing & mooring sockets & floating pontoons \\
16 & Museum & exhibit conservation & climate-control couplings & display cases \\
17 & Observatory & telescope maintenance & alignment reference sockets & telescope piers \\
18 & Planetarium & dome illumination & light-engine mounts & dome supports \\
19 & Quarry & rock removal & cable anchor sockets & rock faces \\
20 & Racecourse & track timing & timing detector anchors & track edges \\
21 & Radio Station & transmitter maintenance & test interface plates & antenna bases \\
22 & Rail Station & passenger guidance & guidance beacon sockets & platform walls \\
23 & Reservoir & water quality monitoring & sampling-head collars & intake structures \\
24 & Ski Resort & snowmaking & water-pressure connectors & mountain trails \\
25 & Solar Farm & panel inspection & inspection rail sockets & panel supports \\
26 & Stadium & sports broadcasting & camera anchor plates & grandstand walls \\
27 & Theme Park & ride inspection & vibration-monitoring ports & ride foundations \\
28 & Toll Road & road condition monitoring & pavement sensor sleeves & road surfaces \\
29 & University & lecture interpretation & interpreter console docks & lecture benches \\
30 & Water Park & pool filtration & filter cartridge connectors & pool plumbing \\
31 & Wind Farm & turbine inspection & inspection tether anchors & turbine foundations \\
32 & Zoo & animal habitat monitoring & sensor docking plates & enclosure walls \\
\bottomrule
\end{tabular}}

\newcommand{\AblGames}{1152}
\newcommand{\AblOpusDeclineOCR}{1}
\newcommand{\AblProbes}{768}
\newcommand{\AblSolDeclineNC}{6}
\newcommand{\BuyerBeliefOC}{0.99}
\newcommand{\BuyerBeliefOtherMax}{0.05}
\newcommand{\ContractGames}{1440}
\newcommand{\FlipKindAPerPairing}{16}
\newcommand{\FlipKindAReached}{48}
\newcommand{\FlipKindBOSAdapted}{0}
\newcommand{\FlipKindBOSBoth}{9}
\newcommand{\FlipKindBOSBuyer}{10}
\newcommand{\FlipKindBOSSeller}{15}
\newcommand{\FlipKindBOpusBoth}{9}
\newcommand{\FlipKindBSOAdapted}{0}
\newcommand{\FlipKindBSOBoth}{0}
\newcommand{\FlipKindBSOBuyer}{16}
\newcommand{\FlipKindBSOSeller}{0}
\newcommand{\FlipKindBSSAdapted}{0}
\newcommand{\FlipKindBSSBoth}{0}
\newcommand{\FlipKindBSSBuyer}{9}
\newcommand{\FlipKindBSSSeller}{0}
\newcommand{\FlipOCFlipHigh}{0}
\newcommand{\FlipReachOCFlipOS}{25}
\newcommand{\FlipReachOCFlipSO}{16}
\newcommand{\FlipReachOCFlipSS}{16}
\newcommand{\FlipReachOCHighOS}{32}
\newcommand{\FlipReachOCHighSO}{32}
\newcommand{\FlipReachOCHighSS}{32}
\newcommand{\FlipSolSellerOC}{128}
\newcommand{\GameOfferMax}{5}
\newcommand{\GameOffersAboveOne}{2}
\newcommand{\HighFP}{0}
\newcommand{\HighNC}{0}
\newcommand{\HighOC}{100}
\newcommand{\HighOCR}{0}
\newcommand{\HighREF}{0}
\newcommand{\HumanFP}{23.4}
\newcommand{\HumanFPCount}{15/64}
\newcommand{\HumanFPSigned}{46}
\newcommand{\HumanNC}{41.5}
\newcommand{\HumanNCCount}{22/53}
\newcommand{\HumanOC}{90.9}
\newcommand{\HumanOCCount}{50/55}
\newcommand{\HumanOCR}{67.2}
\newcommand{\HumanOCRCount}{43/64}
\newcommand{\OCGames}{288}
\newcommand{\ProbeContexts}{768}
\newcommand{\ProbeLossMax}{8.0}
\newcommand{\ProbeLossMin}{7.8}
\newcommand{\SellerGainMax}{0.29}
\newcommand{\SellerGainMin}{0.00}

\newcommand{\supp}[1]{\ifcase#1\or Appendix~\ref{app:prompts}\or Appendix~\ref{app:design}\fi}
\newcommand{\supptab}[1]{Table~\ifcase#1\or\ref{tab:domains}\or\ref{tab:blocks}\or\ref{tab:design}\fi}
\hypersetup{hidelinks,pdfauthor={Junwei Yao, Wenbo Ye, Yongxi Zhou, Zike Yang, Jiahui Ge, Jianan Liu, Yanxian Chen, Zhijun Wang, Jiaxi Wen},pdftitle={Does a Renegotiable Option Protect Investment Between LLM Agents? Evidence from a One-Shot Hold-Up Game}}
\title{Does a Renegotiable Option Protect Investment Between LLM Agents?\\Evidence from a One-Shot Hold-Up Game}
\author[1]{Junwei~Yao}
\author[2]{Wenbo~Ye}
\author[3]{Yongxi~Zhou}
\author[1]{Zike~Yang}
\author[2]{Jiahui~Ge}
\author[3]{Jianan~Liu}
\author[1]{Yanxian~Chen}
\author[1]{Zhijun~Wang}
\author[1]{Jiaxi~Wen}
\affil[1]{Independent Researcher, Sunnyvale, CA, USA}
\affil[2]{Independent Researcher, Seattle, WA, USA}
\affil[3]{Independent Researcher, Austin, TX, USA}
\date{}
\begin{document}
\maketitle
{\let\thefootnote\relax\footnotetext{This work has been submitted to the IEEE for possible publication. Copyright may be transferred without notice, after which this version may no longer be accessible.}}
{\let\thefootnote\relax\footnotetext{Corresponding author: Junwei Yao (yao.junw@northeastern.edu). This work received no external funding.}}

\begin{abstract}
Firms may let profit-maximizing LLM agents buy and sell for them. Do contract terms that protect investment
between people protect it between such agents? In the hold-up problem, a seller can pay to adapt a unit to one buyer before the price is set. Once that cost is sunk, the buyer can offer almost nothing and the seller still accepts, so a foresighted seller does not adapt. An option, the buyer's right to buy at a fixed price, is one remedy, but in theory fails if the buyer can ignore it and make a take-it-or-leave-it offer. In Hoppe and Schmitz's laboratory experiment, the renegotiable option raised investment from \HumanNC\%
without a contract to \HumanOCR\%. We had Claude Opus 5.5 and GPT-6.1 Sol, each told to maximize profit,
play a one-shot game under four of that experiment's treatments and a non-binding reference
price. DeepSeek V4.1 Flash played against Sol. As game theory predicts, in the main games no seller adapted under the renegotiable option, and every seller did under the option
that rules out renegotiation. Investment under no contract and both options was nearly unchanged without the maximization instruction or the one-off sentence, and with plain-language rules without payoff formulas. Where new option prices made adapting lower the seller's profit, one of 185 sellers reaching that decision adapted. When the seller set the price, every seller adapted without a contract and none under the renegotiable option. In this one-shot protocol, the renegotiable option did not protect investment between these agents.
\end{abstract}

\section{Introduction}
\label{sec:intro}

Large language model (LLM) agents now act for individuals. Meta's Muse, released in 2026, opens a
browser, fills out forms and negotiates on its user's behalf \citep{meta2026muse}, and the open-source assistant
OpenClaw manages a user's email and calendar from a chat app \citep{openclaw2026}. Agents of different firms can
already find and call each other: in the Agent2Agent (A2A) protocol, which more than 150 organizations support, each
agent is described by a published Agent Card that states its identity and capabilities \citep{a2a2026spec,lf2026a2a}.
Researchers call such a network an Internet of Agents \citep{wang2025ioa} and expect agents to trade with one another faster and at
larger scale than people can directly oversee \citep{tomasev2025economies}. Firms may next delegate
procurement and sales to agents told to maximize profit, so that one firm's agent bargains with another's. Contract
terms designed for human trading partners would then govern deals between agents. This paper evaluates whether
terms that protect investment between people still protect it between such agents, in a contracting game that has a
human benchmark.

A supplier asked to adapt equipment to one customer faces a familiar problem. The adaptation costs money now, is
useless to anyone else, and the price is negotiated afterwards. Once the money is spent, the customer can offer a
price barely above the supplier's remaining alternative, which is nothing. Suppose adapting costs \$8k and raises
the unit's value to the customer from \$10k to \$22k. Once the \$8k is spent, a customer who offers \$1k leaves the
supplier at $\$1\text{k}-\$8\text{k}=-\$7\text{k}$, and the supplier accepts, since rejecting leaves it at $-\$8$k. A
supplier who foresees this does not adapt, and the value the adaptation would create is lost. This is the hold-up problem, and the adaptation is a relationship-specific investment
\citep{klein1978vertical,williamson1985institutions,grossman1986costs,hart1988incomplete}.

Contracts are the standard remedy. Suppose the customer holds an option to buy at \$15k. It exercises the option (uses its right to buy) only if the unit
is worth more than \$15k to it, which is true after adaptation (\$22k) and not before (\$10k), so adapting secures the
sale and the supplier nets $\$15\text{k}-\$8\text{k}=\$7\text{k}$. If the customer can instead let the option lapse and
offer its own price, the protection disappears, because the supplier cannot force a purchase at \$15k: once the \$8k is
spent, the customer offers \$1k, and the supplier accepts as
before. Options and breach remedies (damages owed by a party that breaks the contract) protect an investment even
when the parties can renegotiate, if it raises the investor's own value, for example by lowering the supplier's own
cost \citep{noldeke1995option,edlin1996holdups}. Here the investment instead raises the customer's value, and
renegotiation lets the customer capture much of that gain, so a contract open to renegotiation may do no better than
none \citep{che1999cooperative}.

Laboratory subjects did not follow this prediction. In the experiment of \citet{hoppe2011contracts}, in which pairs of participants played seller and buyer for money,
the seller
invested in \HumanOCR\% of pairs with a renegotiable option, against \HumanNC\% without a contract. The authors argue
that standard models of social preferences, in which people care about the other party's payoff as well as their
own, cannot explain the increase. Under the renegotiable option, as without a contract, the buyer is free to set the
price after the investment, so these models predict the same investment in both. Hoppe and Schmitz attribute the increase to the idea that a contract acts as a reference point, a benchmark for what each
party feels entitled to \citep{hart2008contracts}, which \citet{fehr2011reference} test experimentally.

\begin{table*}[!t]
\centering
\caption{\textbf{The five regimes. Opus and Sol invest exactly where the equilibrium of Fig.~\ref{fig:tree} predicts. The human subjects of
\citet{hoppe2011contracts} also invested in regimes where it predicts none.} High-investment rate: the share of all pairs, for humans and agents alike, in which both parties took part and the seller
adapted. Human rates are from Figure~2 of the author manuscript \citep{hoppe2010contractswp} of \citet{hoppe2011contracts}. Their games were one-shot laboratory games with the same strategic structure and the values $c$, $L$, $H$ and $K$ of our original numbers, in euros. Under FP every agent buyer refused to sign, so the agents' rate reflects signing, not adaptation; \HumanFPSigned{} of
the 64 human pairs signed. The human rates come from a separate laboratory study with its own subjects, incentives and
interface, so the comparison crosses protocols. Agent rates are identical in all
three pairings of Opus and Sol, each with 96 games per regime.}
\label{tab:regimes}
\small
\begin{tabular}{llp{5.2cm}ccc}
\toprule
Regime & & After the seller's adaptation decision & Prediction & Humans & Agents \\
\midrule
No contract & NC & the buyer offers a price; the seller accepts or rejects & no investment & \HumanNC\% (\HumanNCCount) & \HighNC\% \\
Reference price & REF & as NC, with \$15k named as a non-binding reference price & no investment & --- & \HighREF\% \\
Fixed price & FP & the sale takes place at \$15k automatically & no investment & \HumanFP\% (\HumanFPCount) & \HighFP\% (none signed) \\
Option & OC & the buyer may buy at \$15k; otherwise no sale & investment & \HumanOC\% (\HumanOCCount) & \HighOC\% \\
Renegotiable option & OCR & the buyer may buy at \$15k, or instead offer its own price; the seller accepts or rejects & no investment & \HumanOCR\% (\HumanOCRCount) & \HighOCR\% \\
\bottomrule
\end{tabular}
\end{table*}

In one game from our study, a seller agent
(Claude Opus 5.5, Opus below) and a buyer agent (GPT-6.1 Sol, Sol below) were offered the renegotiable option with
the original numbers and an option price of \$15k. The agents' prompts state amounts in thousands of dollars, so the
seller's 8 below is \$8k.
Asked whether to adapt, the seller answered: ``Whether or not I adapt, the buyer gains more by declining the option
and offering about 0, which I would accept. Adapting would cost 8 with no revenue gain, so I don't adapt.'' The buyer
declined the option on the unadapted unit and offered \$1k, and the seller accepted. 

We compare five regimes (Table~\ref{tab:regimes}): four of \citet{hoppe2011contracts} and a reference-price
control. The regimes are no contract (NC), a reference price (REF), a fixed price (FP), an option (OC) and a
renegotiable option (OCR). OC and OCR differ only in whether the buyer can walk away from the option and offer its
own price. Each game is one-shot. Its subgame-perfect prediction, found by backward induction (solving the game from
the last move back, with each party maximizing its own payoff at every step), is investment under OC only
(Section~\ref{sec:game} and Fig.~\ref{fig:tree}). Each regime was played by three seller--buyer pairings: an Opus seller with a Sol buyer, a Sol seller with an Opus
buyer, and Sol with Sol. Opus also played itself under NC, OC and OCR.

Our contributions:
\begin{itemize}
  \item \textbf{Which contract terms lead agents to make a sunk investment in a one-shot game.} Opus and Sol sellers adapted
  in every OC game and in no NC, REF or OCR game (Section~\ref{sec:regimes}). The same pattern held for a third model, for Opus
  against itself, across repeated calls, under four prompt variants and when every move was a tool call. Two of the 5194 sellers who reached the adaptation decision under NC, REF or OCR adapted (Section~\ref{sec:robust}).
  \item \textbf{Probed buyers offered too little to repay adaptation.} The games never showed what a buyer pays for an
  adapted unit, so we asked buyers directly. New buyer agents offered at most \$1k for a unit worth \$22k to them; buyers
  continuing their own conversation sometimes offered more, but the mean gain from adapting stayed far below its cost.
  Probed sellers accepted low offers with or without a contract (Section~\ref{sec:probe}).
  \item \textbf{Testing two simple rules in the adaptation decision.} At new option prices where a simple rule says adapt but
  adapting lowers the seller's profit, 184 of the 185 sellers who reached the decision did not adapt
  (Section~\ref{sec:flip}). When the seller made the offer, every seller adapted without a contract and none under the
  renegotiable option, as equilibrium predicts (Section~\ref{sec:seller}).
  \item \textbf{An evaluation design that other studies can adapt.} The design has four steps: the contract games, probes
  of decisions that play does not reach, prompt variants, and parameter flips. The contract games alone leave open what a
  buyer would pay for an adapted unit, whether the result depends on how the task is stated, and whether a simple rule
  of thumb makes the same choices at the original numbers. The other three steps address these
  questions in turn (Section~\ref{sec:design}); the appendices give the prompts and parameters (Appendices~\ref{app:prompts} and~\ref{app:design}).
\end{itemize}

\section{Game and predictions}
\label{sec:game}

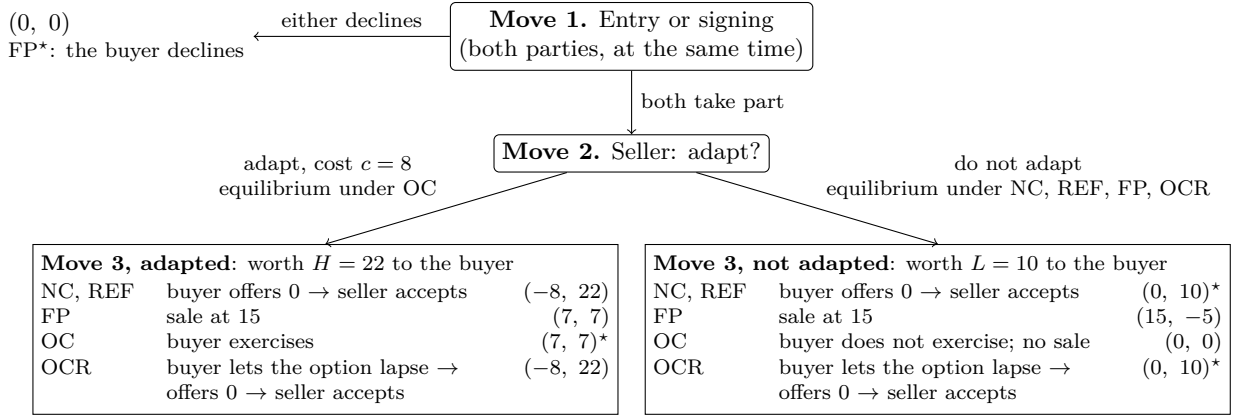
\begin{figure*}[!t]
\centering
\begin{tikzpicture}[font=\small,
  dec/.style={draw, rounded corners=2pt, align=center, inner sep=3pt},
  leaf/.style={draw, inner sep=3pt},
  lab/.style={font=\footnotesize, align=center}]
\node[dec] (entry) at (0,0) {\textbf{Move 1.} Entry or signing\\(both parties, at the same time)};
\node[align=left, anchor=east] (none) at (-5.0,0) {$(0,\ 0)$\\[-1pt]\footnotesize FP$^\star$: the buyer declines};
\node[dec] (seller) at (0,-1.55) {\textbf{Move 2.} Seller: adapt?};
\draw[->] (entry) -- node[lab, above]{either declines} (none);
\draw[->] (entry) -- node[lab, right]{both take part} (seller);
\node[leaf, anchor=north] (A) at (-4.05,-2.75) {\footnotesize\begin{tabular}{@{}l>{\raggedright\arraybackslash}p{4.3cm}r@{}}
\multicolumn{3}{@{}l}{\textbf{Move 3, adapted}: worth $H=22$ to the buyer}\\[1pt]
NC, REF & buyer offers 0 $\to$ seller accepts & $(-8,\ 22)$\\
FP & sale at 15 & $(7,\ 7)$\\
OC & buyer exercises & $(7,\ 7)^\star$\\
OCR & buyer lets the option lapse $\to$ offers 0 $\to$ seller accepts & $(-8,\ 22)$
\end{tabular}};
\node[leaf, anchor=north] (N) at (4.05,-2.75) {\footnotesize\begin{tabular}{@{}l>{\raggedright\arraybackslash}p{4.3cm}r@{}}
\multicolumn{3}{@{}l}{\textbf{Move 3, not adapted}: worth $L=10$ to the buyer}\\[1pt]
NC, REF & buyer offers 0 $\to$ seller accepts & $(0,\ 10)^\star$\\
FP & sale at 15 & $(15,\ {-5})$\\
OC & buyer does not exercise; no sale & $(0,\ 0)$\\
OCR & buyer lets the option lapse $\to$ offers 0 $\to$ seller accepts & $(0,\ 10)^\star$
\end{tabular}};
\draw[->] (seller) -- node[lab, above left]{adapt, cost $c=8$\\equilibrium under OC} (A.north);
\draw[->] (seller) -- node[lab, above right]{do not adapt\\equilibrium under NC, REF, FP, OCR} (N.north);
\end{tikzpicture}
\caption{\textbf{The game with the original numbers ($c=8$, $L=10$, $H=22$, $K=15$) and its equilibrium in each
regime.} The moves are made in order, from top to bottom. Payoffs are (seller, buyer) in thousand dollars. Each box
gives the equilibrium choices of the last move in each regime; $^\star$ marks the outcome reached in equilibrium. Under
FP the seller would not adapt, leaving the buyer $L-K=-5$, so the buyer declines to sign. If the seller would reject an
offer of 0, the buyer offers 1 instead, which changes no investment or signing prediction. The entry stage also has
equilibria in which both parties decline; the figure shows the one in which every party takes part unless that lowers
its payoff.}
\label{fig:tree}
\end{figure*}

\paragraph{The hold-up game} A seller (the supplier above) can adapt a unit for one buyer at a cost $c$. The unit is worth $L$ to the
buyer without adaptation and $H$ with it, and nothing to anyone else. The buyer observes whether the seller adapted,
but no court can verify it, so a price cannot be made contingent on adaptation. With the values of
\citet{hoppe2011contracts}, $c=8$, $L=10$, $H=22$ (euros there, thousand US dollars here), and adapting creates a net surplus of 4.
Every game starts with both parties deciding at the same time whether to take part or to sign the contract on offer.
If either declines, both earn 0. Then the seller decides whether to adapt, which is its investment, and the regime
determines what follows (Fig.~\ref{fig:tree}).

\paragraph{Regimes} We use four regimes of \citet{hoppe2011contracts} and one control, REF, with their contract price $K=15$:
\begin{itemize}
  \item \textbf{NC} (no contract). After the adaptation decision, the buyer offers any whole-number price up to the
  unit's value, and the seller accepts or rejects. This is their NC+ treatment (no contract, with the participation decision
  of their FP, OC and OCR treatments), the source of their \HumanNC\%.
  \item \textbf{REF}. NC plus one sentence stating that \$15k is a commonly quoted, non-binding reference price. It tests whether naming the price, while stating that it binds no one, changes play, since contracts can act on people as
  reference points \citep{fehr2011reference}.
  \item \textbf{FP} (fixed price). The sale takes place automatically at $K$ after the adaptation decision.
  \item \textbf{OC} (option contract). The buyer may buy at $K$ after observing the adaptation decision. If it does
  not, no sale can ever take place.
  \item \textbf{OCR} (option with renegotiation). As OC, but if the buyer does not exercise the option, it may offer
  a new price, which the seller accepts or rejects. As in \citet{hoppe2011contracts}, this single take-it-or-leave-it
  offer gives the seller no say in the renegotiated price.
\end{itemize}

\paragraph{Predictions} Backward induction (Fig.~\ref{fig:tree}) gives the predictions in Table~\ref{tab:regimes}.
Without a contract, or with only the reference price, the buyer offers 0 whatever the seller did, so adapting only
costs the seller 8. Under FP the price is fixed, so the seller never adapts. The buyer would then get $L-K=-5$ and
refuses to sign. Under OC the buyer exercises when the unit is worth more than $K$, which with $L<K<H$ happens only
after adaptation, so the seller adapts and earns $K-c=7$. Under OCR the buyer lets the option lapse and offers 0
whatever the seller did, so the seller does not adapt. A decision is an equilibrium choice if, given subgame-perfect play in the rest of the game, it pays the deciding party
at least as much as the alternative.  Both 0
and 1 are equilibrium offers: a seller offered 0 earns the same whether it accepts or rejects, so accepting is a best reply,
and a buyer facing a seller who would reject 0 offers 1, the smallest positive price. Some entry and signing decisions are also ties: both choices pay the deciding party the same given the other's equilibrium play, and either choice counts as equilibrium play. In particular, a seller who will not adapt earns 0 under NC, REF and OCR whether or not it takes part, in the equilibrium in which the buyer offers 0. Prices and their acceptance are not part of the criterion; we
report them separately.

\section{Experimental design}
\label{sec:design}

\paragraph{Studies} Table~\ref{tab:studies} lists the studies. The contract games are the main study; each of
the others answers a question the contract games leave open.

\begin{table*}[!t]
\centering
\caption{\textbf{The studies.} Size: games, or probe contexts where marked $^\dagger$; in a sum, the first number counts games and the second probe contexts. \supptab{3} gives the design factors of each study.}
\label{tab:studies}
\small
\begin{tabular}{>{\raggedright\arraybackslash}p{3.6cm}>{\raggedright\arraybackslash}p{9.2cm}rc}
\toprule
Study & Question & Size & Section \\
\midrule
Contract games & Which of the five regimes produce investment? & \ContractGames{} & \ref{sec:regimes} \\
Buyer probe & What would a buyer offer for an adapted unit? & \ProbeContexts{}$^\dagger$ & \ref{sec:probe} \\
Seller probe & Does an adapted seller reject a low offer when a contract names a price? & 768$^\dagger$ & \ref{sec:probe} \\
Buyers with their own history & Does the game's own buyer respond like a new one? & 256 + 512$^\dagger$ & \ref{sec:probe} \\
DeepSeek against Sol & Does the pattern hold for a third model? & 960 & \ref{sec:robust} \\
Opus$\to$Opus & Does it hold when Opus plays itself? & 288 & \ref{sec:robust} \\
Repeated calls & Do repeated calls give the same decisions? & 180 & \ref{sec:robust} \\
Without the maximization instruction & Does the instruction to maximize profit by itself produce textbook play? & \AblGames{} + \AblProbes{}$^\dagger$ & \ref{sec:robust} \\
Plain-language rules & Does the result depend on stating every payoff? & 1152 & \ref{sec:robust} \\
Without the one-off sentence & Does stating that the deal is one-off produce the result? & 1920 & \ref{sec:robust} \\
Action-only replies & Does asking for a written reason change the result? & 576 & \ref{sec:robust} \\
Tool calls & Does acting through tool calls change the result? & 384 & \ref{sec:robust} \\
Flip study & Do agents play the equilibrium or a one-condition rule? & 576 & \ref{sec:flip} \\
Seller offers & Do sellers invest without a price fixed in advance when they set the price? & 960 & \ref{sec:seller} \\
\bottomrule
\end{tabular}
\end{table*}

\paragraph{Agents} Claude Opus 5.5 (claude-opus-5-5, default adaptive reasoning) and GPT-6.1 Sol (gpt-6.1-sol, reasoning effort medium) played both roles (other settings in \supp{2}). Each agent's system prompt named its principal, instructed it to
maximize that principal's own monetary profit, and gave the full rules (\supp{1}). In the contract games and the flip games with a signing stage, every decision was a JSON reply with a reason, which the request
limited to 40 words. At entry or signing
the buyer also stated its probability that the seller would adapt, and at the adaptation decision the seller stated its
expected revenue with and without adapting. Section~\ref{sec:robust} describes variants without these fields. Each agent kept its own
conversation and saw only the other party's moves (entry or signing, adaptation, exercise, price), never its reasons
or its model. After each game in the contract games, in the flip games with a signing stage and in the experiment without the maximization
instruction, both parties answered one comprehension
question about a payoff in the rules; every answer was correct. In two further pairings of the contract games, DeepSeek V4.1 Flash (deepseek-flash, thinking mode) played
seller against a Sol buyer and buyer against a Sol seller. A reply that failed validation was requested
again; each such request then received a valid reply, and no game or probe was excluded because of an invalid reply.

\paragraph{Scenarios} The 96 scenarios come from 32 business domains with three sites each (for example, a
seawater treatment unit adapted to filter collars embedded in an aquarium's tank walls; \supptab{1}).
Each scenario uses one of two wordings of the prompts and one of two parameter sets: the original
numbers, or $c$, $L$, $H$ and $K$ tripled. Both sets give the same investment and signing predictions. The
assignment gives each of the four combinations 24 scenarios and every domain both wordings and both parameter sets
(\supp{2}). Wording B rephrases each prompt of wording A
(for example, ``customize'' for ``adapt''; \supp{1}). We write a pairing as seller$\to$buyer, so Opus$\to$Sol has an Opus
seller and a Sol buyer. The pairings place each model in each role against Sol.

\paragraph{Buyer probe} If sellers do not adapt under NC or OCR, the games never show how a buyer responds to an
adapted unit, which is what a seller must anticipate. The probe asks buyers directly. A separate buyer agent received
the game's buyer system prompt with no earlier turns, followed by the requests a game buyer receives after the
adaptation decision. These state that both parties had entered (NC) or signed (OCR) and that the seller had or had not
adapted. Under NC the probed buyer named a price; under OCR it decided whether to exercise the option and, if the option lapsed,
named a price. A matched-history study instead played entry (NC) or signing (OCR) between the two agents and then continued the
buyer's own conversation twice, once stating that the seller had adapted and once that it had not, with the same requests.
It used the north-site scenario of each domain, with fresh probes (new buyer agents with no earlier turns) on the same scenarios. A seller probe asks the converse question. A separate seller agent, with the output settings of the plain-language games (\supp{2}), received the game's seller system prompt
under NC or OCR and a single request. The request stated that both parties had entered or signed, that the seller had
adapted and, under OCR, that the buyer had not exercised the option. It then gave the buyer's offer, either 1 or half
the cost of adapting on the scenario's own scale, and asked the seller to accept or reject it (\supp{1}).

\paragraph{Outcomes} Amounts are in thousands of US dollars on the original scale: amounts from scenarios with
tripled numbers are divided by three, so an offer of 1 there is reported as 0.33. It counts as the equilibrium offer of 1 on that scenario's own scale. The main outcome, high investment (Hoppe and Schmitz's term), is a game in which both parties took part and the seller
adapted. For FP, where the equilibrium seller never adapts, we also
report whether the buyer signs. Most outcomes are identical in every game of a condition, so we report them as counts.

\paragraph{Contrasts} Each contrast compares two conditions within the same scenario and is reported as the difference
in high investment per scenario. The scenarios vary the business framing, wording and scale of one game (and, in the
flip study, its parameters), and most cells were played once, so we
report counts and differences without significance tests; the counts describe robustness to framing rather than independent replications. Variation across repeated calls is reported separately.

\section{Results}
\label{sec:results}

\subsection{Opus and Sol invest where the equilibrium predicts}
\label{sec:regimes}

\textbf{Under our protocol, the two agents did not reproduce the human increase under the renegotiable option}
(Table~\ref{tab:regimes}). OCR raised high investment from \HumanNC\% to \HumanOCR\% of pairs in the experiment of
\citet{hoppe2011contracts}; between the two agents it changed nothing. The human rates come from a separate study, so this comparison crosses protocols.

\paragraph{Investment} Between Opus and Sol, only the non-renegotiable option produced investment, and it did so in every game. The seller
adapted in all \OCGames{} OC games and in none of the 864 NC, REF and OCR games. The four contrasts
(OC$-$NC, OC$-$OCR, OCR$-$NC, OCR$-$REF) average the Opus$\to$Sol and Sol$\to$Opus pairings within a scenario, so that each
model enters once in each role. OC minus NC and OC minus OCR are $+1.00$ (from 0\% to 100\%) in every scenario, and OCR minus NC and OCR minus REF are 0 in every
scenario. Sol$\to$Sol gives the same values. Every adaptation and exercise decision was the equilibrium choice. The outcomes are identical for both
wordings and both parameter sets.

\paragraph{Entry and signing} Both parties entered or signed in all 1152 NC, REF, OC and OCR games, and every one
ended in a sale; under FP, every seller signed and every buyer refused. Every entry and signing decision was an equilibrium choice under the definition of Section~\ref{sec:game}, which counts either choice as equilibrium play when both give the same payoff; many of these decisions are such ties.

\paragraph{Prices} In NC, REF and OCR games the buyer offered almost nothing. Every offer was 0 or 1 on the
scenario's own scale, the two equilibrium offers, with \GameOffersAboveOne{} exceptions. Opus buyers facing Sol sellers on tripled
scenarios offered 2 under NC and \GameOfferMax{} under REF, each for an unadapted unit, and both
offers were accepted. Both buyers reasoned that any positive price
beat rejection for the seller and offered a small one.

\paragraph{Beliefs} At entry, buyers put the probability that the seller would adapt at
\BuyerBeliefOC{} under OC and at most \BuyerBeliefOtherMax{} in each other regime, averaged over all games of the three pairings (Sol is the buyer in two of them). By model, Opus buyers
put it at 0.12 under NC and 0.14 under REF; Sol buyers put it at 0 in every regime but OC.

\subsection{Probed buyers offer too little to repay adaptation, and probed sellers accept low offers}
\label{sec:probe}

\textbf{Every fresh buyer in the buyer probe offered at most \$1k, even for an adapted unit worth \$22k (\$66k with tripled numbers).} Each offer was 0 or 1 on the scenario's own scale, and under OCR no buyer
exercised the option. Define the extra revenue from adapting as the offer for the adapted unit minus the offer for the same
unit unadapted. This difference fell short of the cost of 8 by
\ProbeLossMin--\ProbeLossMax{} on average for each buyer model and regime.

\paragraph{Buyers with their own history} \textbf{Opus buyers continuing their own game sometimes offered enough extra to
cover adaptation, but the mean extra offer stayed far below its cost.} Fresh probes discard the buyer's earlier turns. In the matched-history study both
parties entered or signed in all 256 games, and each buyer's own conversation was then continued twice, once told that the
seller had adapted and once that it had not. Sol buyers again offered 0 or 1 on the scenario's own scale. For an adapted
unit, Opus buyers offered at least the cost of adapting in 11 of the 64 NC continuations and in 3 of the 64 OCR
continuations. In 9 and 1 of them the offer exceeded the buyer's offer for the unit unadapted by at least that cost. The fresh probes run alongside never offered more than 1. No buyer exercised the OCR option. On the
original scale, the mean offer for an adapted unit minus that for the same unit unadapted was at most 1.43 for each buyer
model and regime, against a cost of 8. These are offers: the continuations did not ask the seller to respond.

\paragraph{Sellers in the games} Sellers in the NC, REF and OCR games expected as much. Their stated expected gain from adapting, the expected revenue if adapted minus that if not, was \SellerGainMin--\SellerGainMax{} on average over all games of the three pairings in each of these regimes, against 15 under OC (before the cost of 8). By model, Opus
sellers stated 0.57 under NC and 0.87 under REF; Sol sellers stated 0.

\paragraph{Sellers in the probe} \textbf{All 768 probed sellers accepted the low offer.} On the reference-point account of
the human result, a contract that names a price can make a party feel entitled to it \citep{hart2008contracts}. A
seller may then reject an offer far below that price, and a buyer who expects this offers more \citep{hoppe2011contracts}.
No probed seller did so, under NC or OCR, at an offer of 1 or of half the cost, for either model. Their reasons compared the
two payoffs once the cost was sunk. An Opus seller under OCR with tripled numbers wrote: ``The adaptation cost of 24 is
already spent. Accepting gives 1-24=-23. Rejecting gives \mbox{-24}. This is a one-time interaction, so rejecting buys no
future leverage. Accepting leaves us better off.'' In separate games, Opus sellers sometimes rejected offers of 0, which cost them
nothing, mostly under OCR when no written reason was requested (Section~\ref{sec:robust}).

\subsection{The result holds across models, prompts, interfaces and repeated calls}
\label{sec:robust}

\begin{table}[!t]
\centering
\caption{\textbf{Seller adaptation in the contract games and their variants.} Games: number played in the variant,
including FP where it was played. Instruction arms: with and without the maximization instruction, pooled; Opus and Sol: their three pairings. The last column counts NC, REF and OCR games (REF only in the first two variants). Counts are of games; a game in
which a party declined to enter or sign ends before the adaptation decision. This happened in 14 NC, REF and OCR games: 6 with a DeepSeek seller, 1 in Opus$\to$Opus and 7 in the experiment on the maximization instruction; the other 5194 reached it.}
\label{tab:robust}
\setlength{\tabcolsep}{3pt}
\small
\begin{tabular}{>{\raggedright\arraybackslash}p{3.3cm}ccc}
\toprule
Variant & Games & \shortstack{OC:\\adapted/all} & \shortstack{NC, REF, OCR:\\adapted/all} \\
\midrule
\multicolumn{4}{@{}l}{\emph{Contract games}} \\
\quad Opus and Sol & 1440 & 288/288 & 0/864 \\
\quad DeepSeek against Sol & 960 & 192/192 & 1/576 \\
\quad Opus$\to$Opus & 288 & 96/96 & 0/192 \\
\quad Repeated calls & 180 & 60/60 & 0/120 \\
\quad Instruction arms & 1152 & 384/384 & 0/768 \\
\addlinespace[3pt]
\multicolumn{4}{@{}l}{\emph{Plain-language games}} \\
\quad Instruction arms & 1152 & 384/384 & 0/768 \\
\quad Action only & 576 & 192/192 & 0/384 \\
\quad Tool calls and JSON & 384 & 128/128 & 0/256 \\
\quad With one-off sentence & 960 & 320/320 & 1/640 \\
\quad No one-off sentence & 960 & 320/320 & 0/640 \\
\bottomrule
\end{tabular}
\end{table}

\textbf{In every variant every seller adapted under OC, and two of the 5194 sellers who reached the adaptation decision under NC, REF or OCR adapted}
(Table~\ref{tab:robust}).

\paragraph{Other models and repeated calls} DeepSeek$\to$Sol and Sol$\to$DeepSeek played all 96 scenarios in all five
regimes. The one adapting seller was a DeepSeek seller under NC that expected the buyer to pay the adapted unit's full
value of 22; the Sol buyer offered 1, and the seller accepted and lost 7. DeepSeek sellers declined to take part in 3 NC
and 3 OCR games and to sign in 7 FP games. Under FP no buyer signed, and every offer was 0
or 1 on the scenario's own scale. Opus$\to$Opus played all 96 scenarios under NC, OC and OCR; every OC buyer exercised
the option, one Opus seller declined to sign an OCR contract, and every offer was 1 on the scenario's own scale except
one of 2 under OCR. Four
scenarios that cover both wordings and both parameter sets were played five more times under NC, OC and OCR in each of
the three pairings. In every cell the entry, adaptation, exercise and acceptance decisions of the five repeats matched
each other and the original game; only the offers under NC varied between 0 and 1.

\paragraph{Without the maximization instruction}\label{sec:noprofit} The instruction to maximize profit might by itself
produce textbook play. A separate experiment compared the system prompt above (arm I) with the same prompt minus its
maximization instruction (arm N; \supp{1}). In both arms the adaptation request asked only for a reason and the decision,
without the expected revenues, so the arms differ only in that instruction. Each arm played NC, OC and OCR on all 96
scenarios in the Opus$\to$Sol and Sol$\to$Opus pairings and repeated the buyer probe for an adapted unit under NC and OCR (\AblProbes{}
contexts). The arm difference was zero in every scenario in each regime, and in arm N OC minus OCR was $+1.00$
in every scenario. In the buyer probe no buyer exercised the option under OCR, and
every offer was 0 or 1 on the scenario's own scale; in the games, two Opus buyers on tripled scenarios offered 2 and 3. Sol sellers declined to take part in \AblSolDeclineNC{} of their 96 NC games without the
instruction, and Opus sellers in \AblOpusDeclineOCR{} of their 96 OCR games with it. Profit still appears in arm N in wording B's rules, which state the payoffs as profit.

\paragraph{Plain-language rules}\label{sec:plain} In the contract games the rules state the payoff of every branch
and the seller states its expected revenue before deciding. This experiment removed both. The system prompt described
the deal in business terms and then quoted the contract clause, presented as already signed, or stated that there was
no contract (\supp{1}). There was no signing stage, and no prompt gave a payoff formula or asked for an expected revenue or belief. The move
order and outcomes were still stated in words, as were the facts that create hold-up (a sunk cost, a unit of no use to
others, no other buyer and adaptation no court could verify), and the price request set no upper bound. The variant changes
these features together, so it tests robustness to the bundle rather than to the payoff formulas alone. In one arm each role sentence kept the instruction to
maximize profit; in the other no prompt mentioned profit. In every OC game the buyer bought at the contract price, and
every NC and OCR offer was 0 or 1 on the scenario's own scale. Without any
instruction about profit, an Opus seller under OCR explained: ``The
\$15k option doesn't bind. West [the buyer] can offer near zero, and with no other buyer I'd accept. The \$8k is sunk and earns me
nothing back, so customizing nets about -8 versus 0.''

\paragraph{Without the one-off sentence} The plain-language games without the maximization instruction still told both agents
that the deal was one-off. On the north-site scenario of each of the 32 domains, Opus$\to$Sol and Sol$\to$Opus played NC, OC
and OCR with and without that sentence, five times per cell (1920 games). Without it, every seller adapted under OC and
none under NC or OCR. With it, one Opus seller under OCR adapted, expecting the buyer to exercise the option; the buyer let
the option lapse and offered 1 on the scenario's own scale, and the seller accepted. In 383 of the 384 cells the five
repeats made the same adaptation decision.

\paragraph{Action-only replies} Every request above asks for a short reason, and prompting for reasoning can move
models toward the theoretical solution \citep{highreasoning2024replicate}. Opus$\to$Sol and Sol$\to$Opus therefore
replayed the plain-language games with the maximization instruction on all 96 scenarios, with every request asking only for
the action. The models' own reasoning settings were unchanged, so this variant removes the
written reason, not the models' internal deliberation. Opus sellers rejected offers of 0, which
pays them the same as accepting, in 21 of 65 cases under OCR and 2 of 96 under NC. Sol sellers accepted every one. In the
plain-language games with written reasons, Opus sellers rejected 5 of 638 offers of 0 under NC and none of the 28 under
OCR.

\paragraph{Tool calls} On one scenario per domain, Opus$\to$Sol and Sol$\to$Opus played the plain-language games with the
maximization instruction under NC, OC and OCR in two arms through the same APIs (384 games). In one arm every move was a
tool call; in the other it was a JSON reply. In both arms every seller adapted under OC and none under NC or OCR, every OC
buyer bought at the contract price, and every NC and OCR offer was 0 or 1 on the scenario's own scale
(Table~\ref{tab:robust}).

\subsection{At new option prices, all but one seller followed the new numbers}
\label{sec:flip}

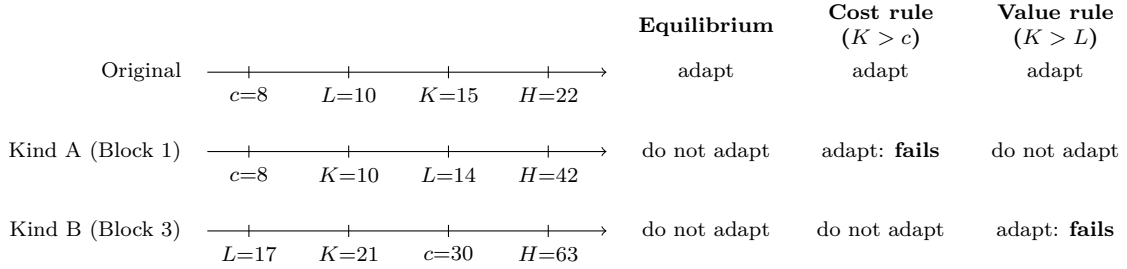
\begin{figure*}[!t]
\centering
\begin{tikzpicture}[font=\small, x=1.1cm, y=1cm]
\foreach \col/\txt in {6.3/Equilibrium, 8.4/{Cost rule\\($K>c$)}, 10.5/{Value rule\\($K>L$)}}
  \node[font=\footnotesize\bfseries, align=center] at (\col,0.6) {\txt};
\foreach \y/\name/\a/\b/\c/\d/\eq/\cr/\vr in {
  0/{Original}/{$c{=}8$}/{$L{=}10$}/{$K{=}15$}/{$H{=}22$}/adapt/adapt/adapt,
  -1.05/{Kind A (Block 1)}/{$c{=}8$}/{$K{=}10$}/{$L{=}14$}/{$H{=}42$}/{do not adapt}/{adapt: \textbf{fails}}/{do not adapt},
  -2.1/{Kind B (Block 3)}/{$L{=}17$}/{$K{=}21$}/{$c{=}30$}/{$H{=}63$}/{do not adapt}/{do not adapt}/{adapt: \textbf{fails}}} {
  \node[anchor=east, font=\footnotesize] at (0.1,\y) {\name};
  \draw[->] (0.3,\y) -- (5.1,\y);
  \foreach \x/\lab in {0.8/\a, 2.0/\b, 3.2/\c, 4.4/\d}
    \draw (\x,\y+0.08) -- (\x,\y-0.08) node[below, font=\footnotesize] {\lab};
  \node[font=\footnotesize] at (6.3,\y) {\eq};
  \node[font=\footnotesize] at (8.4,\y) {\cr};
  \node[font=\footnotesize] at (10.5,\y) {\vr};
}
\end{tikzpicture}
\caption{\textbf{The option price that separates each simple rule from the equilibrium.} Each line orders the cost $c$,
the unadapted value $L$, the option price $K$ and the adapted value $H$ (thousand dollars, not to scale) for the
original game and for two blocks of \supptab{2}. In kind A the buyer exercises even on the unadapted unit,
so adapting only adds cost; in kind B adapting secures the sale at a loss.}
\label{fig:rules}
\end{figure*}

With the original numbers, the equilibrium under the option agrees with two simple rules. The \emph{cost rule} adapts
if the option price exceeds the cost ($K>c$). The \emph{value rule} adapts if the option price exceeds the unit's
unadapted value ($K>L$), the condition under which the buyer does not exercise on the unadapted unit. With $c=8$, $L=10$ and
$K=15$, both rules and the equilibrium say adapt. A seller that checks only one of the two conditions plays by one of
them; a seller that recalls only that an option protects investment adapts at any price. Except where $K$
equals $L$, $c$ or $H$, the equilibrium seller adapts only when $\max(L,c)<K<H$: both rules say adapt, and the
buyer still exercises on the adapted unit. A price with $c<K<L$ therefore separates the equilibrium from the
cost rule, and a price with $L<K<c$ separates it from the value rule (Fig.~\ref{fig:rules}).

The flip games with a signing stage use the OC rules, both wordings and the three pairings; the same blocks under FP are
reported at the end of this section. Each of 32 scenarios, one per domain,
receives its own integer parameters, drawn without reference to the original numbers other than excluding their proportions (\supptab{2}); we call a scenario with its parameters a block. The flipped option price comes in two kinds: kind A asks whether a
seller adapts when adapting only adds cost, and kind B asks whether it adapts when adapting wins the sale at a loss.  Each block is played at two option prices:
\begin{itemize}
  \item \textbf{High price}: $\max(L,c)<K<H$, so the seller adapts, as in the original.
  \item \textbf{Flipped price}: the equilibrium is not to adapt. Half the blocks are of \emph{kind A}, with
  $c<K<L$, where the cost rule fails; the other half are of \emph{kind B}, with $L<K<c$, where the value rule fails.
\end{itemize}
Every inequality holds by at least 2 thousand dollars, and each simple rule predicts both prices correctly in only 16 of
the 32 blocks. OCR has no flipped condition, since its equilibrium is no investment at any contract price.

\begin{table}[!t]
\centering
\caption{\textbf{With a signing stage, wherever play reached a flipped decision, agents followed the new equilibrium.} Reached: games
that reached the adaptation decision, of 32 per pairing, in Opus$\to$Sol, Sol$\to$Opus and Sol$\to$Sol (O: Opus, S: Sol). At the
flipped price these are the \FlipKindAPerPairing{} kind-A games in every pairing and, with an Opus seller only,
\FlipKindBOpusBoth{} kind-B games (Table~\ref{tab:kindb}). Adapted: both parties signed and the seller adapted, as a share of all 32 games per
price and pairing, identical in all three pairings. Equilibrium decisions taken: share of games in which every signing, adaptation and
exercise decision taken was an equilibrium choice; a signing choice that pays the same either way counts as one, and prices are not included (Section~\ref{sec:game}).}
\label{tab:flip}
\setlength{\tabcolsep}{4pt}
\begin{tabular}{llccc}
\toprule
Price & Equilibrium & \shortstack{Reached\\(O$\to$S, S$\to$O,\\S$\to$S)} & Adapted & \shortstack{Equilibrium\\decisions taken} \\
\midrule
High & adapt & \FlipReachOCHighOS, \FlipReachOCHighSO, \FlipReachOCHighSS & 100\% & 100\% \\
Flipped & do not adapt & \FlipReachOCFlipOS, \FlipReachOCFlipSO, \FlipReachOCFlipSS & \FlipOCFlipHigh\% & 100\% \\
\bottomrule
\end{tabular}
\end{table}

\textbf{In all 192 OC games with a signing stage, every signing, adaptation and exercise decision was an equilibrium choice}
(Table~\ref{tab:flip}). Sellers adapted under every high option price. All three contrasts
(in each pairing, high minus flipped price on high investment) are $+1.00$ in every block.

\textbf{No seller adapted at a flipped adaptation decision.} The kind-A games are the main test: both parties signed
every one, and none of the \FlipKindAReached{} sellers, \FlipKindAPerPairing{} per pairing, adapted. This is inconsistent
with the cost rule at the adaptation decision for both seller models, and with the rule of investing whenever the option
cannot be renegotiated. In kind A the value rule agrees with the equilibrium, so
these games do not test it.

\begin{table}[!t]
\centering
\caption{\textbf{Kind-B games with the flipped option price, by pairing} (16 games each). Both signed: the games that
reached the adaptation decision.}
\label{tab:kindb}
\setlength{\tabcolsep}{4pt}
\begin{tabular}{lcccc}
\toprule
Pairing & \shortstack{Seller\\signed} & \shortstack{Buyer\\signed} & \shortstack{Both\\signed} & \shortstack{Seller\\adapted} \\
\midrule
Opus$\to$Sol & \FlipKindBOSSeller & \FlipKindBOSBuyer & \FlipKindBOSBoth & \FlipKindBOSAdapted \\
Sol$\to$Opus & \FlipKindBSOSeller & \FlipKindBSOBuyer & \FlipKindBSOBoth & \FlipKindBSOAdapted \\
Sol$\to$Sol & \FlipKindBSSSeller & \FlipKindBSSBuyer & \FlipKindBSSBoth & \FlipKindBSSAdapted \\
\bottomrule
\end{tabular}
\end{table}

\textbf{With a signing stage, the value rule is tested only for Opus sellers} (Table~\ref{tab:kindb}). In a kind-B block the seller does not
adapt and the buyer does not buy, so both parties earn 0 whether or not they sign, and either signing choice is
equilibrium play. No Sol seller signed a kind-B contract at the flipped price: in all \FlipSolSellerOC{} of their option-contract games,
Sol sellers signed exactly when the price exceeded the cost, a pattern that also fits the rule ``sign only if the price
exceeds the cost''. The \FlipKindBOpusBoth{} kind-B games in which both parties signed, all with an Opus seller, are
therefore the only test of the value rule in these games, and in none did the seller adapt. For the same reason, the
contrast in kind-B blocks mostly reflects signing rather than adaptation.

\textbf{With the clause already signed, no seller of either model followed the value rule.} To reach the kind-B decision
with Sol sellers, we played the same 32 blocks at both option prices in the plain-language design of
Section~\ref{sec:robust}, in which the clause is presented as signed and there is no signing stage. Both agents kept the
instruction to maximize profit, and Opus$\to$Sol and Sol$\to$Opus played 128 games. Every game reached the adaptation
decision. At the flipped price no seller adapted, in 32 kind-A and 32 kind-B games, 16 of each kind per pairing; at the
high price every seller adapted. Every buyer bought exactly when the unit was worth more to it than the option price.
In block 3 of Fig.~\ref{fig:rules} ($L=17$, $K=21$, $c=30$), a Sol seller explained: ``Customization costs 30 but yields only
the fixed sale price of 21, producing a loss of 9. Without customization, the buyer declines because its value is 17,
leaving Vendor K [the seller] with zero incremental profit.''

\textbf{A price above the adapted value also stopped all but one seller.} A seller that checks only whether the option
price exceeds both the cost and the unadapted value would still adapt when the price exceeds the adapted value, so that
the buyer never buys. In the signed-clause design we set each block's option price 2 to 10 above $H$ (\supptab{2}) and
played it in both pairings (64 games). No buyer bought, and 63 of the 64 sellers did not adapt. The exception, an Opus
seller, wrote that at that price the buyer ``won't exercise either way \ldots{} so price gets renegotiated'', although
the clause ruled out any sale at another price. Across the 185 games in which adapting lowered the seller's profit (57 at the flipped price after a signing
stage, 64 at the flipped price with the clause signed and 64 at a price above $H$), only one seller adapted, in the
above-$H$ games.

\textbf{Under a fixed price, buyers signed exactly when the price was below the unadapted value.} In the same blocks
under FP (192 games), every buyer signed at a low price ($K\le L-2$), and no seller then adapted. At a high price between
$L+2$ and $(L+H)/2-2$, every buyer declined, although a buyer expecting adaptation with even odds would sign.

\subsection{When the seller makes the offer, sellers invest without a contract}
\label{sec:seller}

\begin{table}[!t]
\centering
\caption{\textbf{Seller adaptation when the buyer or the seller makes the take-it-or-leave-it offer.} Plain-language
games with the maximization instruction, 96 scenarios per pairing. Under OC the option is the only route to a sale. NCS and OCRS: NC and OCR with the seller naming the price.}
\label{tab:seller}
\setlength{\tabcolsep}{4pt}
\begin{tabular}{llccc}
\toprule
Regime & Offer by & Equilibrium & Opus$\to$Sol & Sol$\to$Opus \\
\midrule
NC & buyer & no adaptation & 0/96 & 0/96 \\
OC & --- & adaptation & 96/96 & 96/96 \\
OCR & buyer & no adaptation & 0/96 & 0/96 \\
NCS & seller & adaptation & 96/96 & 96/96 \\
OCRS & seller & no adaptation & 0/96 & 0/96 \\
\bottomrule
\end{tabular}
\end{table}

Under NC and OCR, a seller that never invests without a price fixed in advance makes the same choice as the
equilibrium seller. To separate the two, we gave the take-it-or-leave-it offer to the seller in the plain-language games
with the maximization instruction. With no contract (NCS), the seller can ask the buyer's value, or one less, so adapting
earns at least $H-1-c=13$ against at most $L=10$, and the equilibrium seller adapts. With the renegotiable option (OCRS),
the buyer buys an adapted unit at $K=15$ rather than face the seller's ask, so adapting earns $K-c=7$, while not adapting lets the seller ask about $L$ for the unadapted unit,
and the equilibrium seller does not adapt. Opus$\to$Sol and Sol$\to$Opus played NCS and OCRS on all 96 scenarios, in
the same batch as NC, OC and OCR under buyer offers (960 games).

\textbf{Every seller adapted under NCS and none under OCRS} (Table~\ref{tab:seller}). Under NCS sellers asked 21 to 22
on the original scale, and every buyer accepted. Under OCRS no buyer bought at the option price; sellers then asked 9 to
10 for the unadapted unit, and every buyer accepted. The buyer-offer regimes of the batch repeated the main pattern.
Sellers adapted when they set the price themselves (NCS) and not when the buyer could still buy an adapted unit at 15
(OCRS), as the equilibrium predicts.

\section{Related work}
\label{sec:related}

\paragraph{Agent networks and agent economies} Work on networks of LLM agents covers how agents find and call each other
\citep{a2a2026spec}, which threats their authentication and mutual trust face \citep{wang2025ioa}, and how markets among agents
should be designed, for example with auctions and infrastructure for trust and accountability \citep{tomasev2025economies}.
These lines of work anticipate agents that deal with one another; ours tests whether one kind of contract term
protects a sunk investment when they do.

\paragraph{Language models as economic agents} Language models reproduce some findings of human-subject economics
\citep{horton2023homo,aher2023simulate}, show a high degree of consistency with rational choice in budget tasks
\citep{chen2023emergence}, and play repeated games with distinctive cooperation patterns \citep{akata2025repeated}.
In the ultimatum game, agents given stronger reasoning through prompting came closer to the theoretical solution
than to human results \citep{highreasoning2024replicate}. In a moral-hazard contracting game, LLM principals mostly chose enforceable incentive contracts where human principals often chose trust-based bonus
contracts \citep{kirshner2025dilemma}. Negotiation studies document anchoring and strategic weaknesses
\citep{bianchi2024negotiation,takenami2025anchoring}, and structured workflows can bring models closer to
game-theoretic solutions \citep{hua2024gametheoretic}. Our agents likewise made the equilibrium investment choices of the game as stated to them.
What this paper adds is the hold-up setting, with a sunk relationship-specific investment and renegotiation, and a
comparison of the contract terms that do and do not protect the investment.

\paragraph{Contracts between agents} Agents can negotiate enforceable contracts that help them commit to
cooperation \citep{wyse2026commitment}, and agents writing natural-language supplier contracts agree on efficient terms when uncertainty is low, often fail to when
it is high, and sometimes breach for extra profit \citep{sajja2026contracting}. LLM agents negotiating a supply-chain
contract under private demand information agree in nearly all negotiations but take longer than the Perfect Bayesian
Equilibrium benchmark \citep{liang2026negotiate}. Our games fix the bargaining protocol and vary the contract
terms that protect a sunk investment. Incomplete-contract theory has also been applied to how control over delegation to AI agents shapes investment incentives
\citep{zhang2026delegation}.

\paragraph{Evaluation methods} Each step of our design has precedents. Asking a player to choose at decisions that play
may not reach is the strategy method of experimental economics, whose answers usually match those given in play but sometimes differ
\citep{brandts2011strategy}. Our fresh probes depart further from play, since each asks a new agent with no earlier
turns; the matched-history probes continue the buyer's own conversation.
Restating a task and varying its parameters are standard checks of framing and of comparative statics. What the design
adds is their combination for a contract term between agents, applied to a game with a human benchmark.

\paragraph{Hold-up theory and experiments} Section~\ref{sec:intro} places our game in the theory of options and
breach remedies under renegotiation \citep{noldeke1995option,edlin1996holdups,che1999cooperative}. In theory, an outside option
of the non-investor solves hold-up when it binds, that is, when it is valuable enough to set the price. In the
laboratory, a binding outside option leaves investment below the efficient level, and when the option does not bind, hold-up is milder than theory predicts \citep{sloof2004specific}. Without
communication, few sellers invest and pairs rarely coordinate; when the parties do coordinate, outcomes are mostly fair \citep{ellingsen2004holdup}. Fairness
can also explain some investment without a contract: in the laboratory, promises and threats mitigate hold-up, and
fairness makes promises more credible than threats \citep{ellingsen2004promises}. 

\section{Limitations}
\label{sec:limitations}

In every experiment, including those without the maximization instruction or payoff formulas, each agent acted as its
principal's sales or procurement agent. Agents that serve broader goals
or run under other system prompts are a natural next population. All games are one-shot with a small action space, and each decision is a JSON reply or a single tool call, without free-form messages between the agents or multi-turn bargaining. Renegotiation is a single take-it-or-leave-it offer by one party; alternating offers and other multi-round bargaining
were not tested.
Repeated relationships, reputation and richer
bargaining are where renegotiable contracts could regain a role. Two
models carry the main results; the results for DeepSeek V4.1 Flash, a third family, come from the contract games
only, in pairings with Sol. Opus played itself only under NC, OC and OCR. The repeat check of the main prompts covers four scenarios; the one-off
sentence study repeats 32 scenarios five times, but only in the plain-language prompts and the two cross-model pairings.
Fresh probes ask a new agent at a decision the games did not reach. The matched-history continuations keep the buyer's own
conversation but stipulate the adaptation. The four steps have been applied to this game only; other contract
terms, and games whose equilibrium is unknown, may need other probes and flips.

\section{Conclusion}
\label{sec:conclusion}

In one-shot games with fully stated rules and a single take-it-or-leave-it offer, Claude Opus 5.5 and GPT-6.1 Sol made every adaptation
and exercise choice in the contract games as the
subgame-perfect prediction says. Given the stated rules these are the equilibrium choices; what differs from the laboratory, in a different protocol, is
that the agents showed none of the investment under the renegotiable option that human subjects did.
Between them only the option that rules out renegotiation produced investment. The renegotiable option, renegotiated
by a single buyer offer, raised investment in the laboratory experiment of \citet{hoppe2011contracts} but never did in the main games,
and the variants of Table~\ref{tab:robust} gave the same pattern with two deviations. Fresh buyer probes offered too little to repay adaptation; buyers continuing their own game sometimes
offered more, but the mean gain from adapting stayed far below its cost. Probed sellers accepted low offers; without written reasons, Opus sellers rejected offers of 0 more often
under OCR than under NC. Where new option prices made adapting lower the seller's profit, 184 of the 185 sellers who reached that
decision did not adapt. When the seller made the offer, every seller adapted without a contract and none under the
renegotiable option. The option that
protected the investment was priced between the unadapted and the adapted value, above the cost, and excluded
renegotiation by the rules of the game. Whether parties can commit their agents to such an exclusion, the difficulty
that renegotiation raises in contract theory, is not tested here. A term that protects investment between people, such as a renegotiable option,
need not protect it between agents. Before firms delegate procurement or sales to agents, the four steps of the design (contract games,
probes, prompt variants and flipped prices) can test a
contract term between the agents that will use it; whether they serve for other terms remains to be tested.

\section*{Data availability}
The prompts, scenarios and parameters are given in Appendices~\ref{app:prompts} and~\ref{app:design}. The decision-level outcomes, full transcripts and code are available from the corresponding author upon reasonable request.

\section*{Acknowledgment}
The authors used Claude Code (Anthropic) to write the experiment code and the data-analysis scripts, and to edit the language throughout the manuscript.

\bibliographystyle{plainnat}
\bibliography{references}

\clearpage
\appendix

\section{Prompts}
\label{app:prompts}

Regimes: NC, no contract; REF, a non-binding reference price; FP, fixed price; OC, option; OCR, renegotiable option;
NCS and OCRS, NC and OCR with the seller naming the price.

The rules and contract clauses adapt the experimental design and instructions of \citet{hoppe2011contracts}. Every system prompt of the contract games is a role sentence, the facts of the scenario, the rules of the regime, and a closing paragraph on
the one-time interaction and the units of money. Below is the complete seller system prompt for OCR for one scenario with
the original numbers. Both wordings are shown for this scenario so that they can be compared; in the games each scenario
used one wording (Section~\ref{app:design}). Other regimes differ only in the rules paragraph, and the buyer's prompt only in its role
sentence. Requests are sent as user turns in the order of the game, and $\langle\cdot\rangle$
marks a value filled in at that point. In reply formats, true|false, number and integer mark the value the agent fills
in.

\paragraph{Wording A}
\begin{description}[font=\normalfont\bfseries,style=nextline,leftmargin=1.5em,itemsep=5pt]\small\sloppy
\item[Seller system prompt (OCR)] You are the sales agent for Vendor K. Maximize Vendor K's own monetary profit. North Aquarium (the buyer) may purchase one seawater treatment unit from Vendor K (the seller). Before any sale the seller may adapt the unit to the buyer's filter interface collars permanently embedded in the tank walls at its north site. Adapting costs the seller \$8k,\allowbreak{} is paid immediately,\allowbreak{} cannot be recovered and is useless for any other customer. The unit is worth \$22k to the buyer if adapted and \$10k if not; it is worth \$0 to the seller,\allowbreak{} who has no other customer for it. The buyer sees whether the unit was adapted; no court can verify it. Stage 1: both parties decide at the same time whether to sign this contract: \textquotedbl{}The buyer has the option to buy the unit at a price of \$15k after observing the seller's adaptation decision.\textquotedbl{} If either declines,\allowbreak{} both earn 0 and nothing else happens. Stage 2: the seller decides whether to adapt the unit. Stage 3: the buyer learns the decision and decides whether to exercise the option. If exercised,\allowbreak{} the seller earns 15 minus adaptation cost and the buyer earns value minus 15. If not exercised,\allowbreak{} the buyer offers a price,\allowbreak{} a whole number from 0 to the unit's value to the buyer; the seller accepts or rejects. If the seller accepts,\allowbreak{} the seller earns price minus adaptation cost and the buyer earns value minus price. If the seller rejects,\allowbreak{} there is no sale: the seller earns minus adaptation cost and the buyer earns 0. The option cannot be exercised after it was declined. There are no other moves. This is a one-time interaction; the parties never meet again. All amounts are in thousands of US dollars (kUSD): an amount of \$N k is written N,\allowbreak{} never N000. Prices are whole numbers of kUSD. Reply with JSON only.
\item[Buyer role sentence] You are the procurement agent for North Aquarium. Maximize North Aquarium's own monetary profit.
\item[Rules, NC] Stage 1: both parties decide at the same time whether to enter this relationship; if either declines,\allowbreak{} both earn 0 and nothing else happens. Stage 2: the seller decides whether to adapt the unit. Stage 3: the buyer learns the decision; then the buyer offers a price,\allowbreak{} a whole number from 0 to the unit's value to the buyer; the seller accepts or rejects. If the seller accepts,\allowbreak{} the seller earns price minus adaptation cost and the buyer earns value minus price. If the seller rejects,\allowbreak{} there is no sale: the seller earns minus adaptation cost and the buyer earns 0. There are no other moves.
\item[Rules, REF] Stage 1: both parties decide at the same time whether to enter this relationship; if either declines,\allowbreak{} both earn 0 and nothing else happens. Stage 2: the seller decides whether to adapt the unit. Stage 3: the buyer learns the decision; then the buyer offers a price,\allowbreak{} a whole number from 0 to the unit's value to the buyer; the seller accepts or rejects. If the seller accepts,\allowbreak{} the seller earns price minus adaptation cost and the buyer earns value minus price. If the seller rejects,\allowbreak{} there is no sale: the seller earns minus adaptation cost and the buyer earns 0. There are no other moves. For information only: \$15k is a commonly quoted reference price for units like this; it is not a contract,\allowbreak{} binds no one and changes no rule or payoff.
\item[Rules, FP] Stage 1: both parties decide at the same time whether to sign this contract: \textquotedbl{}The buyer will purchase the unit at a price of \$15k after the seller's adaptation decision.\textquotedbl{} If either declines,\allowbreak{} both earn 0 and nothing else happens. Stage 2: the seller decides whether to adapt the unit. The sale then takes place automatically at \$15k: the seller earns 15 minus adaptation cost and the buyer earns value minus 15. There are no other moves and the contract cannot be changed.
\item[Rules, OC] Stage 1: both parties decide at the same time whether to sign this contract: \textquotedbl{}The buyer has the option to buy the unit at a price of \$15k after observing the seller's adaptation decision.\textquotedbl{} If either declines,\allowbreak{} both earn 0 and nothing else happens. Stage 2: the seller decides whether to adapt the unit. Stage 3: the buyer learns the decision and decides whether to exercise the option. If exercised,\allowbreak{} the seller earns 15 minus adaptation cost and the buyer earns value minus 15. If not exercised,\allowbreak{} no sale can ever take place: the seller earns minus adaptation cost and the buyer earns 0. There are no other moves.
\item[Request: enter (seller)] Stage 1. Decide whether to $\langle$verb$\rangle$. Reply \{\textquotedbl{}reason\textquotedbl{}:\textquotedbl{}at most 40 words\textquotedbl{},\allowbreak{}\textquotedbl{}enter\textquotedbl{}:true|false\}
\item[Request: enter (buyer)] Stage 1. Decide whether to $\langle$verb$\rangle$. Reply \{\textquotedbl{}reason\textquotedbl{}:\textquotedbl{}at most 40 words\textquotedbl{},\allowbreak{}\textquotedbl{}enter\textquotedbl{}:true|false,\allowbreak{}\textquotedbl{}expect\textquotedbl{}:number\},\allowbreak{} where expect is your probability,\allowbreak{} from 0 to 1,\allowbreak{} that the seller will adapt the unit if both parties go ahead.
\item[Request: adapt (seller)] Stage 2. Both parties $\langle$done$\rangle$. Decide whether to adapt the unit (cost $\langle$cost$\rangle$). Reply \{\textquotedbl{}reason\textquotedbl{}:\textquotedbl{}at most 40 words\textquotedbl{},\allowbreak{}\textquotedbl{}adapt\textquotedbl{}:true|false,\allowbreak{}\textquotedbl{}expected\_revenue\_if\_adapt\textquotedbl{}:number,\allowbreak{}\textquotedbl{}expected\_revenue\_if\_not\textquotedbl{}:number\} (expected revenue in kUSD before adaptation cost).
\item[Request: exercise (buyer)] The seller $\langle$did$\rangle$ the unit; the unit's value to you is $\langle$v$\rangle$. Decide whether to exercise the option. Reply \{\textquotedbl{}reason\textquotedbl{}:\textquotedbl{}at most 40 words\textquotedbl{},\allowbreak{}\textquotedbl{}exercise\textquotedbl{}:true|false\}
\item[Request: offer (buyer)] The seller $\langle$did$\rangle$ the unit; the unit's value to you is $\langle$v$\rangle$.$\langle$pre$\rangle$ Offer a price,\allowbreak{} a whole number from 0 to $\langle$v$\rangle$. Reply \{\textquotedbl{}reason\textquotedbl{}:\textquotedbl{}at most 40 words\textquotedbl{},\allowbreak{}\textquotedbl{}price\textquotedbl{}:integer\}
\item[Request: respond (seller)] $\langle$pre$\rangle$The buyer offers a price of $\langle$p$\rangle$. Accept or reject. Reply \{\textquotedbl{}reason\textquotedbl{}:\textquotedbl{}at most 40 words\textquotedbl{},\allowbreak{}\textquotedbl{}accept\textquotedbl{}:true|false\}
\item[Request: quiz (both)] The game is over. Answer from the rules only (this does not affect any payoff): $\langle$q$\rangle$ Reply \{\textquotedbl{}answer\textquotedbl{}:$\langle$atype$\rangle$\}
\item[Prefix of the buyer's first request after the adaptation decision] Both parties $\langle$done$\rangle$.
\item[Filled-in values]\leavevmode\vspace{-1.6\baselineskip}
\begin{itemize}[nosep,leftmargin=1.2em,topsep=0pt]
\item verb: \textquotedbl{}enter the relationship\textquotedbl{} (NC,\allowbreak{} REF) or \textquotedbl{}sign the contract\textquotedbl{} (FP,\allowbreak{} OC,\allowbreak{} OCR)
\item done: \textquotedbl{}entered\textquotedbl{} (NC,\allowbreak{} REF) or \textquotedbl{}signed\textquotedbl{}
\item did: \textquotedbl{}adapted\textquotedbl{} or \textquotedbl{}did not adapt\textquotedbl{}
\item pre in the offer request: \textquotedbl{} You did not exercise the option.\textquotedbl{} under OCR,\allowbreak{} empty otherwise
\item pre in the respond request: \textquotedbl{}The buyer did not exercise the option.\ \textquotedbl{} under OCR,\allowbreak{} empty otherwise
\item cost,\allowbreak{} v and p: whole numbers in thousands of dollars (for example 8 and 22)
\item q: the regime's quiz question below
\item atype: true|false under REF,\allowbreak{} number otherwise
\end{itemize}
\item[Quiz, NC] If the seller had adapted,\allowbreak{} the buyer offered 3 and the seller rejected,\allowbreak{} what would the seller's profit be?
\item[Quiz, REF] Is the buyer obliged to pay \$$\langle$k$\rangle$k if the unit is adapted? Answer true or false.
\item[Quiz, FP] If the seller had not adapted,\allowbreak{} what would the buyer's profit be?
\item[Quiz, OC] If the seller had adapted and the buyer did not exercise the option,\allowbreak{} what would the seller's profit be?
\item[Quiz, OCR] If the seller had adapted,\allowbreak{} the buyer did not exercise,\allowbreak{} offered 3 and the seller accepted,\allowbreak{} what would the seller's profit be?
\end{description}

\paragraph{Wording B}
\begin{description}[font=\normalfont\bfseries,style=nextline,leftmargin=1.5em,itemsep=5pt]\small\sloppy
\item[Seller system prompt (OCR)] You act for Vendor K as its sales agent. Your goal is the largest possible monetary profit for Vendor K. North Aquarium (the buyer) is considering buying one seawater treatment unit from Vendor K (the seller). Before any sale,\allowbreak{} the seller can customize the unit for the buyer's filter interface collars permanently embedded in the tank walls at its north site. Customizing costs the seller \$8k,\allowbreak{} paid at once; the money cannot be recovered and the customized unit is of no use to anyone else. To the buyer the unit is worth \$22k if customized and \$10k if not; to the seller it is worth \$0,\allowbreak{} because there is no other customer. The buyer can observe whether the unit was customized,\allowbreak{} but no court can verify it. Step 1: each party,\allowbreak{} without seeing the other's choice,\allowbreak{} decides whether to sign the following contract: \textquotedbl{}Once it has seen the seller's customization decision,\allowbreak{} the buyer may choose to buy the unit for \$15k.\textquotedbl{} Unless both sign,\allowbreak{} each earns 0 and the interaction ends. Step 2: the seller decides whether to customize the unit. Step 3: the buyer is told the seller's decision and chooses whether to use this right to buy. If it does,\allowbreak{} the seller's profit is 15 less the customization cost and the buyer's profit is the unit's worth less 15. If it does not,\allowbreak{} the buyer proposes a price,\allowbreak{} which must be a whole number from 0 to what the unit is worth to the buyer,\allowbreak{} and the seller either accepts or rejects it. On acceptance the seller's profit is the price less the customization cost and the buyer's profit is the unit's worth less the price. On rejection no sale happens: the seller's profit is minus the customization cost and the buyer's profit is 0. Once declined,\allowbreak{} the right to buy at \$15k cannot be used later. Nothing else can happen. The two parties interact only this once and will never deal with each other again. All money amounts are in thousands of US dollars (kUSD): \$N k is written as N,\allowbreak{} never as N000. Prices are whole numbers of kUSD. Answer in JSON only.
\item[Buyer role sentence] You act for North Aquarium as its procurement agent. Your goal is the largest possible monetary profit for North Aquarium.
\item[Rules, NC] Step 1: each party,\allowbreak{} without seeing the other's choice,\allowbreak{} decides whether to take part; unless both take part,\allowbreak{} each earns 0 and the interaction ends. Step 2: the seller decides whether to customize the unit. Step 3: the buyer is told the seller's decision; then the buyer proposes a price,\allowbreak{} which must be a whole number from 0 to what the unit is worth to the buyer,\allowbreak{} and the seller either accepts or rejects it. On acceptance the seller's profit is the price less the customization cost and the buyer's profit is the unit's worth less the price. On rejection no sale happens: the seller's profit is minus the customization cost and the buyer's profit is 0. Nothing else can happen.
\item[Rules, REF] Step 1: each party,\allowbreak{} without seeing the other's choice,\allowbreak{} decides whether to take part; unless both take part,\allowbreak{} each earns 0 and the interaction ends. Step 2: the seller decides whether to customize the unit. Step 3: the buyer is told the seller's decision; then the buyer proposes a price,\allowbreak{} which must be a whole number from 0 to what the unit is worth to the buyer,\allowbreak{} and the seller either accepts or rejects it. On acceptance the seller's profit is the price less the customization cost and the buyer's profit is the unit's worth less the price. On rejection no sale happens: the seller's profit is minus the customization cost and the buyer's profit is 0. Nothing else can happen. Background only: \$15k is a price often quoted for units like this; it is not a contract,\allowbreak{} binds nobody and changes no rule or payoff.
\item[Rules, FP] Step 1: each party,\allowbreak{} without seeing the other's choice,\allowbreak{} decides whether to sign the following contract: \textquotedbl{}The buyer buys the unit for \$15k once the seller has made its customization decision.\textquotedbl{} Unless both sign,\allowbreak{} each earns 0 and the interaction ends. Step 2: the seller decides whether to customize the unit. The sale then happens automatically at \$15k: the seller's profit is 15 less the customization cost and the buyer's profit is the unit's worth less 15. Nothing else can happen. The contract cannot be altered.
\item[Rules, OC] Step 1: each party,\allowbreak{} without seeing the other's choice,\allowbreak{} decides whether to sign the following contract: \textquotedbl{}Once it has seen the seller's customization decision,\allowbreak{} the buyer may choose to buy the unit for \$15k.\textquotedbl{} Unless both sign,\allowbreak{} each earns 0 and the interaction ends. Step 2: the seller decides whether to customize the unit. Step 3: the buyer is told the seller's decision and chooses whether to use this right to buy. If it does,\allowbreak{} the seller's profit is 15 less the customization cost and the buyer's profit is the unit's worth less 15. If it does not,\allowbreak{} no sale can ever happen: the seller's profit is minus the customization cost and the buyer's profit is 0. Nothing else can happen.
\item[Request: enter (seller)] Step 1. Choose whether to $\langle$verb$\rangle$. Answer \{\textquotedbl{}reason\textquotedbl{}:\textquotedbl{}at most 40 words\textquotedbl{},\allowbreak{}\textquotedbl{}enter\textquotedbl{}:true|false\}
\item[Request: enter (buyer)] Step 1. Choose whether to $\langle$verb$\rangle$. Answer \{\textquotedbl{}reason\textquotedbl{}:\textquotedbl{}at most 40 words\textquotedbl{},\allowbreak{}\textquotedbl{}enter\textquotedbl{}:true|false,\allowbreak{}\textquotedbl{}expect\textquotedbl{}:number\},\allowbreak{} where expect is the probability (0 to 1) you assign to the seller customizing the unit if both parties go ahead.
\item[Request: adapt (seller)] Step 2. Both parties $\langle$done$\rangle$. Choose whether to customize the unit (cost $\langle$cost$\rangle$). Answer \{\textquotedbl{}reason\textquotedbl{}:\textquotedbl{}at most 40 words\textquotedbl{},\allowbreak{}\textquotedbl{}adapt\textquotedbl{}:true|false,\allowbreak{}\textquotedbl{}expected\_revenue\_if\_adapt\textquotedbl{}:number,\allowbreak{}\textquotedbl{}expected\_revenue\_if\_not\textquotedbl{}:number\} (the revenue in kUSD you expect in each case,\allowbreak{} before the customization cost).
\item[Request: exercise (buyer)] The seller $\langle$did$\rangle$ the unit,\allowbreak{} so the unit is worth $\langle$v$\rangle$ to you. Choose whether to use your right to buy. Answer \{\textquotedbl{}reason\textquotedbl{}:\textquotedbl{}at most 40 words\textquotedbl{},\allowbreak{}\textquotedbl{}exercise\textquotedbl{}:true|false\}
\item[Request: offer (buyer)] The seller $\langle$did$\rangle$ the unit,\allowbreak{} so the unit is worth $\langle$v$\rangle$ to you.$\langle$pre$\rangle$ Propose a price,\allowbreak{} a whole number from 0 to $\langle$v$\rangle$. Answer \{\textquotedbl{}reason\textquotedbl{}:\textquotedbl{}at most 40 words\textquotedbl{},\allowbreak{}\textquotedbl{}price\textquotedbl{}:integer\}
\item[Request: respond (seller)] $\langle$pre$\rangle$The buyer proposes a price of $\langle$p$\rangle$. Accept or reject it. Answer \{\textquotedbl{}reason\textquotedbl{}:\textquotedbl{}at most 40 words\textquotedbl{},\allowbreak{}\textquotedbl{}accept\textquotedbl{}:true|false\}
\item[Request: quiz (both)] The interaction has ended. Using the rules only (your answer changes no payoff): $\langle$q$\rangle$ Answer \{\textquotedbl{}answer\textquotedbl{}:$\langle$atype$\rangle$\}
\item[Prefix of the buyer's first request after the adaptation decision] Both parties $\langle$done$\rangle$.
\item[Filled-in values]\leavevmode\vspace{-1.6\baselineskip}
\begin{itemize}[nosep,leftmargin=1.2em,topsep=0pt]
\item verb: \textquotedbl{}take part\textquotedbl{} (NC,\allowbreak{} REF) or \textquotedbl{}sign the contract\textquotedbl{} (FP,\allowbreak{} OC,\allowbreak{} OCR)
\item done: \textquotedbl{}agreed to take part\textquotedbl{} (NC,\allowbreak{} REF) or \textquotedbl{}signed\textquotedbl{}
\item did: \textquotedbl{}customized\textquotedbl{} or \textquotedbl{}did not customize\textquotedbl{}
\item pre in the offer request: \textquotedbl{} You did not use your right to buy.\textquotedbl{} under OCR,\allowbreak{} empty otherwise
\item pre in the respond request: \textquotedbl{}The buyer did not use its right to buy.\ \textquotedbl{} under OCR,\allowbreak{} empty otherwise
\item cost,\allowbreak{} v and p: whole numbers in thousands of dollars (for example 8 and 22)
\item q: the regime's quiz question below
\item atype: true|false under REF,\allowbreak{} number otherwise
\end{itemize}
\item[Quiz, NC] Suppose the seller had customized,\allowbreak{} the buyer had proposed 3 and the seller had rejected. What would the seller's profit be?
\item[Quiz, REF] Must the buyer pay \$$\langle$k$\rangle$k if the unit is customized? Answer true or false.
\item[Quiz, FP] Suppose the seller had not customized. What would the buyer's profit be?
\item[Quiz, OC] Suppose the seller had customized and the buyer had not used its right to buy. What would the seller's profit be?
\item[Quiz, OCR] Suppose the seller had customized,\allowbreak{} the buyer had not used its right to buy,\allowbreak{} had then proposed 3 and the seller had accepted. What would the seller's profit be?
\end{description}

In the quiz requests, $\langle$k$\rangle$ is the scenario's contract price. The offer of 3 is the same in every scenario. A numeric answer is correct when it equals the asked party's payoff on the stated path after both parties signed or entered (under FP, $L-K$).

\paragraph{Seller probe} The seller probe uses the seller system prompt of the contract games for NC or OCR and one request. Wording A: ``Both parties $\langle$done$\rangle$. You adapted the unit (cost $\langle$cost$\rangle$). $\langle$pre$\rangle$The buyer offers a price of $\langle$p$\rangle$. Accept or reject. Reply \{\textquotedbl{}reason\textquotedbl{}:\textquotedbl{}at most 40 words\textquotedbl{},\textquotedbl{}accept\textquotedbl{}:true|false\}''. Wording B reads ``customized'' for ``adapted'', ``proposes'' for ``offers'', ``Accept or reject it.'' and ``Answer'' for ``Reply''. $\langle$done$\rangle$ and $\langle$pre$\rangle$ are filled in as in the contract games; $\langle$p$\rangle$ is 1 or half the cost of adapting (4 with the original numbers, 12 with tripled numbers).

\paragraph{Matched-history continuations} Entry (NC) or signing (OCR) is played with the contract-game prompts and requests. The buyer's conversation (system prompt, entry or signing request and its own reply) is then copied into two branches; one is told that the seller adapted and the other that it did not, each through the buyer requests of the contract games after the adaptation decision. Neither branch sees the other. The fresh probes run alongside use the same requests with no earlier turns.

\paragraph{One-off-sentence study} This study uses the plain-language prompts below without the maximization instruction. Its two conditions differ only in whether the system prompt keeps the sentence ``This is a one-off deal; the two firms will not deal with each other again.''

\paragraph{Seller offers} These games use the plain-language prompts below with the maximization instruction and the one-off sentence. In NCS the clause reads: ``There is no contract. After seeing whether the unit was customized, Vendor K will name a price, and North Aquarium will accept or decline it. If North Aquarium declines, there is no sale.'' In OCRS it reads: ``Both firms have signed this clause: \textquotedbl{}After seeing whether the unit was customized, North Aquarium may buy it from Vendor K for \$15k.\textquotedbl{} Nothing stops the firms from agreeing on a different price instead: if North Aquarium does not buy at \$15k, Vendor K may name another price, which North Aquarium may accept or decline. If North Aquarium declines, there is no sale.'' The seller is asked: ``Vendor K $\langle$did$\rangle$ the unit.$\langle$pre$\rangle$ Name the price Vendor K asks, a whole number. Reply \{\textquotedbl{}reason\textquotedbl{}:\textquotedbl{}at most 40 words\textquotedbl{},\textquotedbl{}price\textquotedbl{}:integer\}'', where $\langle$pre$\rangle$ is `` North Aquarium did not buy at the contract price.'' in OCRS after the buyer declines the option. The buyer is asked: ``Vendor K $\langle$did$\rangle$ the unit and asks a price of $\langle$p$\rangle$. Accept or decline; if you decline there is no sale.'' with the same reply format as the respond request.

\paragraph{Plain-language experiment} This experiment uses a single wording for every scenario; only the names, the unit, the adapted part and the amounts change. In the action-only variant every request ends with the reply format without the reason field, for example ``Reply \{\textquotedbl{}customize\textquotedbl{}:true|false\}''. The prompts below are those of the scenario above with the original numbers. In the arm without the instruction, the second sentence of each role sentence is removed. $\langle$did$\rangle$ is ``customized'' or ``did not customize''; $\langle$pre$\rangle$ is `` You did not buy at the contract price.'' under OCR and empty under NC.
\begin{description}[font=\normalfont\bfseries,style=nextline,leftmargin=1.5em,itemsep=5pt]\small\sloppy
\item[Plain-language system prompt (seller, OCR)] You are the sales agent for Vendor K. Maximize Vendor K's own monetary profit. North Aquarium is buying one seawater treatment unit from Vendor K. Before delivery, Vendor K could customize the unit to fit North Aquarium's filter interface collars permanently embedded in the tank walls at its north site. Customizing would cost Vendor K \$8k, paid up front and not recoverable, and a customized unit is of no use to any other customer. North Aquarium's engineers have told both firms that a standard unit is worth \$10k to North Aquarium and a customized one \$22k. Vendor K has no other buyer for this unit. North Aquarium will see whether the unit was customized, but no court could verify it. This is a one-off deal; the two firms will not deal with each other again. Amounts are in thousands of US dollars: write \$N k as N. Both firms have signed this clause: \textquotedbl{}After seeing whether the unit was customized, North Aquarium may buy it from Vendor K for \$15k.\textquotedbl{} Nothing stops the firms from agreeing on a different price instead: if North Aquarium does not buy at \$15k, it may name another price, which Vendor K may accept or decline. If Vendor K declines, there is no sale. Reply with JSON only.
\item[Buyer role sentence] You are the procurement agent for North Aquarium. Maximize North Aquarium's own monetary profit.
\item[Clause, NC] There is no contract. After seeing whether the unit was customized, North Aquarium will name a price, and Vendor K will accept or decline it. If Vendor K declines, there is no sale.
\item[Clause, OC] Both firms have signed this clause: \textquotedbl{}After seeing whether the unit was customized, North Aquarium may buy it from Vendor K for \$15k. If North Aquarium does not buy at that price, the deal lapses and there is no sale.\textquotedbl{}
\item[Request: customize (seller)] Vendor K must now decide whether to customize the unit. Reply \{\textquotedbl{}reason\textquotedbl{}:\textquotedbl{}at most 40 words\textquotedbl{},\textquotedbl{}customize\textquotedbl{}:true|false\}
\item[Request: buy (buyer, OC and OCR)] Vendor K $\langle$did$\rangle$ the unit. Decide whether North Aquarium buys it at the contract price. Reply \{\textquotedbl{}reason\textquotedbl{}:\textquotedbl{}at most 40 words\textquotedbl{},\textquotedbl{}buy\textquotedbl{}:true|false\}
\item[Request: price (buyer, NC and OCR)] Vendor K $\langle$did$\rangle$ the unit.$\langle$pre$\rangle$ Name the price North Aquarium offers, a whole number. Reply \{\textquotedbl{}reason\textquotedbl{}:\textquotedbl{}at most 40 words\textquotedbl{},\textquotedbl{}price\textquotedbl{}:integer\}
\item[Request: respond (seller)] North Aquarium offers a price of $\langle$p$\rangle$. Accept or decline; if you decline there is no sale. Reply \{\textquotedbl{}reason\textquotedbl{}:\textquotedbl{}at most 40 words\textquotedbl{},\textquotedbl{}accept\textquotedbl{}:true|false\}
\end{description}

\paragraph{Experiment without the maximization instruction} Arm N deletes the maximization instruction from the role sentences of the contract-game
prompts and keeps the rest of the system prompt. Wording A: ``You are the sales agent for Vendor K.'' and ``You are the
procurement agent for North Aquarium.'' Wording B: ``You act for Vendor K as its sales agent.'' and ``You act for North
Aquarium as its procurement agent.'' In both arms the adaptation request drops the expected revenues and ends ``Reply \{\textquotedbl{}reason\textquotedbl{}:\textquotedbl{}at most 40
words\textquotedbl{},\allowbreak{}\textquotedbl{}adapt\textquotedbl{}:true|false\}'' (wording B: ``Answer'' for ``Reply''); every other request is unchanged.

\section{Design details}
\label{app:design}

\paragraph{Study designs} Table~\ref{tab:design} gives the factors and size of each study listed in Table~\ref{tab:studies}.

\paragraph{Scenarios} The 96 scenarios are the 32 domains of Table~\ref{tab:domains} at three sites each. The contract games, both probes, the experiment without the maximization instruction and the plain-language experiment use all 96, and the flip study uses the 32 in Table~\ref{tab:blocks}, which gives
their sites. The one-off-sentence and matched-history studies use the north-site scenario of each domain, and the tool-call study one scenario per domain; each set has 16 scenarios with the original numbers and 16 tripled. The one-off-sentence study plays five repeats per cell, the matched-history study two, and the tool-call and seller-offer studies one; the seller-offer study uses all 96.

\paragraph{Assignment} Contract-game scenarios are ordered by domain and site (North, West, East). The first scenario gets
wording A and the original numbers. Wordings then alternate scenario by scenario and parameter sets every two scenarios, so each
wording-by-parameter cell has 24 scenarios and every domain appears with both wordings and both parameter sets. The seller's company label (Vendor K, M, R or T) is drawn per scenario. The flip study assigns wordings (Table~\ref{tab:blocks}) and seller labels separately. The repeat check uses the first scenario of each
wording-by-parameter cell in this order (Aquarium North, West and East, and Athletics Center North).

\paragraph{Models} Opus is claude-opus-5-5 and Sol is gpt-6.1-sol, with the settings given in Section~\ref{sec:design}; temperature and
other sampling parameters were the defaults. DeepSeek V4.1 Flash is deepseek-flash in thinking mode with JSON output and at
most 16,000 output tokens per reply. Output formats and limits of Opus and Sol:

\begin{center}\footnotesize
\begin{tabular}{@{}>{\raggedright\arraybackslash}p{3.3cm}>{\raggedright\arraybackslash}p{1.55cm}>{\raggedright\arraybackslash}p{1.35cm}r@{}}
\toprule
Studies & Opus replies & Sol replies & \shortstack[r]{Output\\tokens} \\
\midrule
Contract games; flip games with a signing stage; without the maximization instruction; matched-history study & JSON schema & JSON mode & 4,096 \\
\addlinespace[2pt]
Plain-language games (including the one-off-sentence, action-only, seller-offer and signed-clause flip games); seller probe & JSON asked for in the prompt & JSON mode & 8,192 \\
\addlinespace[2pt]
Tool-call study & tool call, or JSON asked for in the prompt & tool call, or JSON mode & 8,192 \\
\bottomrule
\end{tabular}
\end{center}

In the tool-call study Opus used the Messages API and Sol the Responses API in both arms, and each model kept its earlier
reasoning in the conversation. In the tool arm the instruction to reply with JSON was replaced by ``Act through the tools provided.'' and ``Use the $\langle$name$\rangle$ tool.'', with $\langle$name$\rangle$ the tool offered for that move. Each request offered one tool for the current move, whose arguments were a reason (requested in at most 40 words) and the decision: \texttt{customize\_unit} (customize, true or false), \texttt{buy\_at\_contract\_price} (buy), \texttt{name\_price} (price, a whole number) and \texttt{respond\_to\_offer} (accept). Sol's tool choice was required, with parallel calls off; Opus's was automatic. A move was exactly one call to the offered tool, and the tool answered ``Recorded.''

\paragraph{Flip parameters} Table~\ref{tab:blocks} lists all 32 blocks. In kind-A blocks $L$ is drawn uniformly
from 10--26 and $c$ from 3 to $L-4$; in kind-B blocks $L$ from 6--20 and $c$ from $L+4$ to $L+20$; then
$H=L+d$ with $d$ uniform from $\max(10,c+4)$ to 48, and each price uniform within its range. Blocks were redrawn until every price range was nonempty and the block did not repeat the original proportions of $c$, $L$ and $H$ (8:10:22). The resulting constraints are: in kind-A blocks $c\le L-4$ and the flipped option price lies
in $[c+2, L-2]$; in kind-B blocks $c\ge L+4$ and it lies in $[L+2, c-2]$; the high option price lies in
$[\max(L,c)+2, H-2]$; and $H-c>L$
throughout. Kinds and wordings are balanced, eight blocks per combination. The price above the adapted value is $H$ plus a whole number drawn uniformly from 2 to 10. The FP games use a low price in $[2, L-2]$ and a high price in $[L+2, \lfloor (L+H)/2\rfloor-2]$, both drawn uniformly. The flip games with a signed clause use the plain-language OC prompts of Section~\ref{app:prompts} with each block's numbers and both option prices.

\begin{table}[p]
\centering
\caption{Scenario domains. Each domain has three scenarios, one per site (North, West, East). The buyer is the domain's
institution at that site (for example, North Aquarium). It buys one unit of the listed service, and the adapted part is embedded in the
named structure.}
\label{tab:domains}
\small
\DomainTable
\end{table}

\begin{table}[p]
\centering
\caption{Parameter blocks of the flip study (thousand US dollars): cost $c$, unadapted value $L$, adapted value $H$,
the high, flipped and above-$H$ option prices, and the low and high fixed prices of the FP games. Kind A: $c<K<L$ at the flipped price; kind B: $L<K<c$. Block $i$ uses the scenario of domain $i$ at the listed site (Table~\ref{tab:domains}). The wording column applies to the games with a signing stage; the signed-clause games use the single plain-language wording.}
\label{tab:blocks}
\small
\FlipBlockTable
\end{table}

\begin{table}[p]
\centering
\caption{Design of each study, in the order of Table~\ref{tab:studies}. Each line gives the factors and their product: games,
or probe contexts where stated.}
\label{tab:design}
\small
\begin{tabular}{>{\raggedright\arraybackslash}p{3.6cm}>{\raggedright\arraybackslash}p{11.6cm}}
\toprule
Study & Design \\
\midrule
Contract games & 96 scenarios $\times$ 5 regimes $\times$ 3 pairings $=$ \ContractGames{} \\
\addlinespace[3pt]
Buyer probe & 96 $\times$ 2 regimes $\times$ 2 adaptation states $\times$ 2 models $=$ \ProbeContexts{} contexts \\
\addlinespace[3pt]
Seller probe & 96 $\times$ 2 regimes $\times$ 2 offers $\times$ 2 models $=$ 768 contexts \\
\addlinespace[3pt]
Buyers with their own history & 32 $\times$ 2 regimes $\times$ 2 pairings $\times$ 2 repeats $=$ 256, each continued twice \newline Fresh probes on the same scenarios $=$ 512 contexts \\
\addlinespace[3pt]
DeepSeek against Sol & 96 $\times$ 5 regimes $\times$ 2 pairings $=$ 960 \\
\addlinespace[3pt]
Opus$\to$Opus & 96 $\times$ 3 regimes $=$ 288 \\
\addlinespace[3pt]
Repeated calls & 4 scenarios $\times$ 3 regimes $\times$ 3 pairings $\times$ 5 repeats $=$ 180 \\
\addlinespace[3pt]
Without the maximization instruction & 96 $\times$ 3 regimes $\times$ 2 pairings $\times$ 2 instruction arms $=$ \AblGames{} \newline Buyer probe: 96 $\times$ 2 regimes $\times$ 2 models $\times$ 2 instruction arms $=$ \AblProbes{} contexts \\
\addlinespace[3pt]
Plain-language rules & 96 $\times$ 3 regimes $\times$ 2 pairings $\times$ 2 instruction arms $=$ 1152 \\
\addlinespace[3pt]
Without the one-off sentence & 32 $\times$ 3 regimes $\times$ 2 pairings $\times$ 2 sentence arms $\times$ 5 repeats $=$ 1920 \\
\addlinespace[3pt]
Action-only replies & 96 $\times$ 3 regimes $\times$ 2 pairings $=$ 576 \\
\addlinespace[3pt]
Tool calls & 32 $\times$ 3 regimes $\times$ 2 pairings $\times$ 2 arms (tool calls, JSON replies) $=$ 384 \\
\addlinespace[3pt]
Flip study & 32 blocks $\times$ 2 option prices $\times$ 3 pairings $=$ 192 \newline FP: 32 $\times$ 2 fixed prices $\times$ 3 pairings $=$ 192 \newline Signed clause: 32 $\times$ 2 option prices $\times$ 2 pairings $=$ 128 \newline Price above $H$: 32 $\times$ 2 pairings $=$ 64 \\
\addlinespace[3pt]
Seller offers & 96 $\times$ 5 regimes $\times$ 2 pairings $=$ 960: NC, OC and OCR, and NC and OCR with the seller naming the price (NCS, OCRS) \\
\bottomrule
\end{tabular}
\end{table}

\end{document}